# The Age and Erosion Rate of Young Sedimentary Rock on Mars

An Y. Li[1,2] , Edwin S. Kite[1] , and Katarina Keating[1,3]
[1] University of Chicago, Department of the Geophysical Sciences, 5734 S. Ellis Avenue Chicago, IL 60637, USA; anli7@uw.edu
[2] University of Washington, Department of Earth and Space Sciences, 4000 15th Avenue, NE Seattle, WA 98195-1310, USA
[3] University of Michigan, Department of Earth and Environmental Sciences, 1100 North University Avenue, Ann Arbor, MI 48109-1005, USA

## Abstract

The Medusae Fossae Formation (MFF) is an enigmatic sedimentary unit near the equator of Mars, with an uncertain formation process and absolute age. Due to the heavily wind-eroded surface, it is difficult to determine the absolute model age of the MFF using a one-parameter model based on the crater size–frequency distribution function with existing crater count data. We create a new two-parameter model that estimates both age and a constant erosion rate ($\beta$) by treating cratering as a random Poisson process. Our study uses new crater count data collected from Context Camera imagery for both the MFF and other young equatorial sedimentary rock. Based on our new model, the Central MFF formed >1.5 Gyr ago and had low erosion rates (<650 nm yr$^{-1}$), whereas the East MFF, Far East MFF, and Zephyria Planum most likely formed <1.5 Gyr ago and had higher erosion rates (>740 nm yr$^{-1}$). The top of Aeolis Mons (informally known as Mount Sharp) in Gale Crater and Eastern Candor have relatively young ages and low erosion rates. Based on the estimated erosion rates (since fast erosion permits metastable shallow ice), we also identify several sites, including Zephyria Planum, as plausible locations for shallow subsurface equatorial water ice that is detectable by gamma-ray spectroscopy or neutron spectroscopy. In addition to confirming <1.5 Gyr sedimentary rock formations on Mars, and distinguishing older and younger MFF sites, we find that fast-eroding locations have younger ages and MFF locations with slower erosion have older best-fit ages.

*Unified Astronomy Thesaurus concepts:* Mars (1007); Surface processes (2116); Planetary surfaces (2113)

## 1. Introduction

The Medusae Fossae Formation (MFF) is a large sedimentary unit located near the equator of Mars (130°–240°E, 15°S–15°N). As one of the largest sedimentary deposits on Mars, the MFF spans an area of $2.1 \times 10^6$ km$^2$ and has a volume of over $1.4 \times 10^6$ km$^3$ (Bradley et al. 2002), yet its formation process, composition, and absolute age are all uncertain (Figure 1; Scott & Tanaka 1986; Bradley et al. 2002; Kerber & Head 2010). The deposits are fine-grained, friable, and easily eroded by the wind (Schultz & Lutz 1988; Tanaka 2000). Due to the MFF's highly erodible nature, its original area (Kerber & Head 2010; Wasilewski & Gregg 2021) and volume may have been double their present-day values (Tanaka 2000; Bradley et al. 2002), so the erosional history of the MFF is important to consider, alongside its age of formation. The MFF deposits may be volcaniclastic (Zimbelman et al. 1997; Mandt et al. 2008; Kerber & Head 2010; Ojha & Lewis 2018; Ojha et al. 2019), although rhythmite has also been reported (Kite et al. 2015), which would likely require liquid water instead (Lewis et al. 2008), so their origin remains a topic of debate. The MFF resembles the upper part of Aeolis Mons, informally known as Mount Sharp, in Gale Crater (Thomson et al. 2008; Thomson et al. 2011; Zimbelman & Scheidt 2012; Wang et al. 2018). Sounding radar, neutron spectroscopy (NS), gamma-ray spectroscopy (GRS), and radar surface reflectivity data suggest that the MFF has shallow ice, but theory predicts shallow ice to be unstable at the equator of Mars, and the question remains open (Watters et al. 2007; Carter et al. 2009; Ojha & Lewis 2018; Pathare et al. 2018; Wilson et al. 2018; Mouginis-Mark & Zimbelman 2020; Campbell et al. 2021).

Stratigraphic studies suggest that the MFF initially formed in the Hesperian, but was later partially reworked during the Amazonian (Kerber & Head 2010; Kerber et al. 2011). A young age of the MFF would suggest Hesperian/Amazonian aqueous cementation as lithifying the sediment, which would imply (if it occurred) that some recent liquid water had to exist at the equator of Mars. This could correspond to small amounts of water over brief periods (Kite et al. 2013; Koeppel et al. 2022). Thus, the age of the MFF could be an important constraint on models of the surface liquid water availability on Mars. Due to the heavily wind-eroded surface, it is difficult to estimate the absolute age of the MFF using existing crater count data and a one-parameter model based on the crater size–frequency distribution (CSFD; Tanaka 2000; Kerber & Head 2010). With the exception of Palucis et al. (2020), existing crater chronology methods are typically not capable of solving simultaneously for erosion and finite age. As a result, a new pipeline is needed that takes into account both age and erosion rate in order to estimate the age of the MFF.

It is expected that crater counts would underestimate the formation age of the MFF, since there are repeated periods of burial interrupted by exhumation (Schultz & Lutz 1988; Schultz 2002). Because both burial and erosion obliterate craters, crater counts would lead to an estimation of the modification age, rather than only the erosion rate (Greeley et al. 2001; Kerber & Head 2010). However, our model finds a simultaneous fit for both erosion rate ($\beta$) and age. This dual approach is more nuanced than only fitting for age, which ignores the underabundance of small craters, as shown in Figures 2 and 3. Figure 2 shows that age alone does not match the crater count data when we compare different best-fit age estimates with negligible ($\beta = 1$ nm yr$^{-1}$) rates of erosion. None of the three fits without erosion agree with the real data set, as they all overestimate the number of craters in smaller crater size bins, in contrast with our new model's fit (this improved agreement is quantified in Section 4.3). As described in Figure 3, this is due to the preferential erosion of smaller craters over time, which is known as the Öpik effect (Öpik 1966; Kite & Mayer 2017 also provide a quantitative example). Because of the limitations of the one-parameter approach, previous studies frequently either fit age up to only the largest craters on young sedimentary rock or they avoid this type of terrain altogether (Platz et al. 2013).

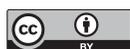

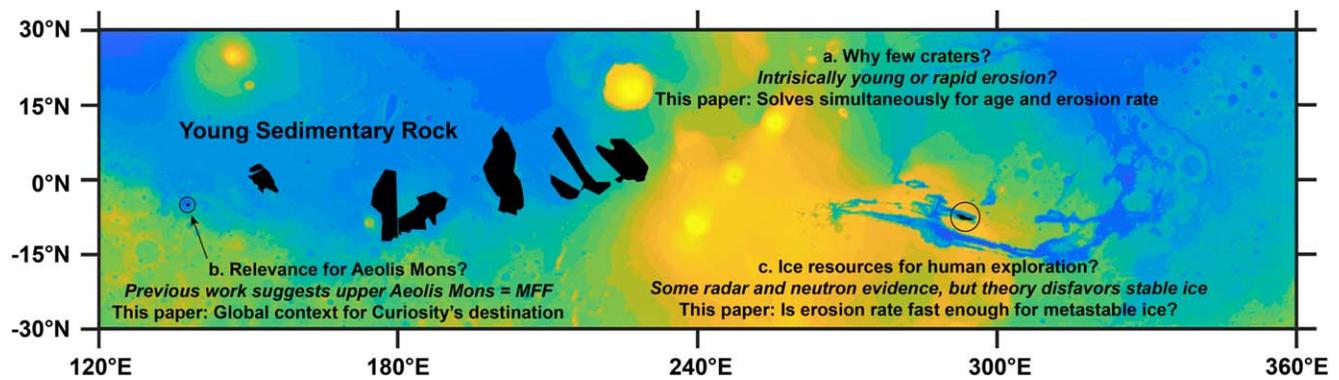

**Figure 1.** This paper's new model addresses questions about the MFF's paucity of craters as related to age, connection to the Curiosity rover field site, and potential for subsurface ice. The black polygons are the sites that are studied in this paper. (a) While crater counts suggest a young formation age for the MFF, in the Middle to Late Amazonian (Scott & Tanaka 1986; Greeley & Guest 1987; Werner 2008), some stratigraphic relations also point to Hesperian ages (Bradley & Sakimoto 2001; Bradley et al. 2002; Kerber & Head 2010). The MFF may have undergone regular reworking and re-induration (Kerber et al. 2011). (b) As discussed more at the end of Section 4.5, the upper part of Aeolis Mons (known informally as Mount Sharp) in Gale Crater and the MFF have similar material and geological features (Thomson et al. 2008; Thomson et al. 2011; Zimbelman & Scheidt 2012; Wang et al. 2018) and similar scales of rhythmic bedding (Lewis & Aharonson 2014), suggesting a similar formation mechanism (e.g., Zimbelman & Griffin 2010; Thomson et al. 2011; Lewis & Aharonson 2014). As the Curiosity rover is headed toward this location, we might have a chance to validate our age and erosion rate estimations. (c) An active subject of debate, some interpretations of the MFF using radar and NS data have suggested the presence of water ice in the subsurface (e.g., Watters et al. 2007; Mandt et al. 2008; Carter et al. 2009; Ojha & Lewis 2018; Wilson et al. 2018; Campbell et al. 2021). The darkest blue color in this figure corresponds to an elevation of −6 km and the brightest yellow corresponds to an elevation of +22 km.

**Table 1**
Area, Location, and Number of Craters for Each Site

| USGS Planetary Nomenclature | Name | Site | Area (km²) | Geographic Location (°E,°N) | Number of Craters |
|---|---|---|---|---|---|
| Memnonia Sulci in Lucus Planum | Central Medusae Fossae 1 A | *1a* | $2.8 \times 10^4$ | −174, −1 | 112 |
| Memnonia Sulci in Lucus Planum | Central Medusae Fossae 1 B | *1b* | $8.8 \times 10^4$ | −177, −8 | 138 |
| Memnonia Sulci in Lucus Planum | Central Medusae Fossae 1 C | *1c* | $1.1 \times 10^5$ | −173, −4 | 147 |
| Zephyria Planum | Zephyria Planum | *2* | $7.8 \times 10^4$ | 153, 0 | 36 |
| Juventae Mensa in Candor Chasma | Eastern Candor | *3* | $6.7 \times 10^3$ | −67, −8 | 5 |
| Lucus Planum | Central Medusae Fossae 2 A | *4a* | $6.4 \times 10^4$ | 179, −9 | 141 |
| Lucus Planum | Central Medusae Fossae 2 B | *4b* | $5.3 \times 10^4$ | 178, −1 | 94 |
| Lucus Planum | Central Medusae Fossae 2 C | *4c* | $9.8 \times 10^4$ | 177, −5 | 944 |
| Amazonis Mensa | Far East Medusae Fossae A | *5a* | $5.5 \times 10^4$ | −147, −2 | 12 |
| Gordii Dorsum | Far East Medusae Fossae B | *5b* | $1.3 \times 10^5$ | −144, 3 | 151 |
| Gordii Dorsum | Far East Medusae Fossae C | *5c* | $2.0 \times 10^5$ | −134, 3 | 101 |
| Eumenides Dorsum | East Medusae Fossae A | *6a* | $1.2 \times 10^5$ | −160, 8 | 153 |
| Eumenides Dorsum | East Medusae Fossae B | *6b* | $2.3 \times 10^5$ | −158, −1 | 80 |
| Aeolis Mons | Upper Mount Sharp | *7* | $1.0 \times 10^3$ | 138, −5 | 5 |
| | All Sites Aggregated | *All* | $1.3 \times 10^6$ | −140, 3 | 2119 |

In this study, to better quantify the age and erosion rate of all young sedimentary rock near the equator of Mars, using new crater count data, we examined 14 sites and applied a probabilistic model that accounts for both erosion and age. Our model assumes that there is one relatively rapid pulse of sedimentation, with no hiatuses, before erosion then occurs at a constant rate. Since there is evidence of multiple cycles of erosion and deposition, along with varying rates of erosion, the model proposed in this study is a simple next step appropriate for the small number of craters available in our data set (see Table 1). Simultaneous fitting for both erosion and age has previously been attempted by Palucis et al. (2020), who used a Monte Carlo approach to subsample parent surfaces generated over a range of ages and β values in order to find a corresponding CSFD. The study site was then assigned the most likely surface ages and β values when the model generated a count of craters that matched up to 25% of the observed crater count in all size bins (Palucis et al. 2020). Our approach is different, since we (a) treat cratering as a random Poisson process in order to apply a probabilistic model to find the best-fit age and erosion rate together, (b) use a new crater data set, focusing on apparently younger sedimentary rock in the equatorial regions of Mars, (c) combine our new data with our new model to infer regional trends between age and erosion rate, and (d) apply our erosion rate estimates to test the claims of subsurface ice.

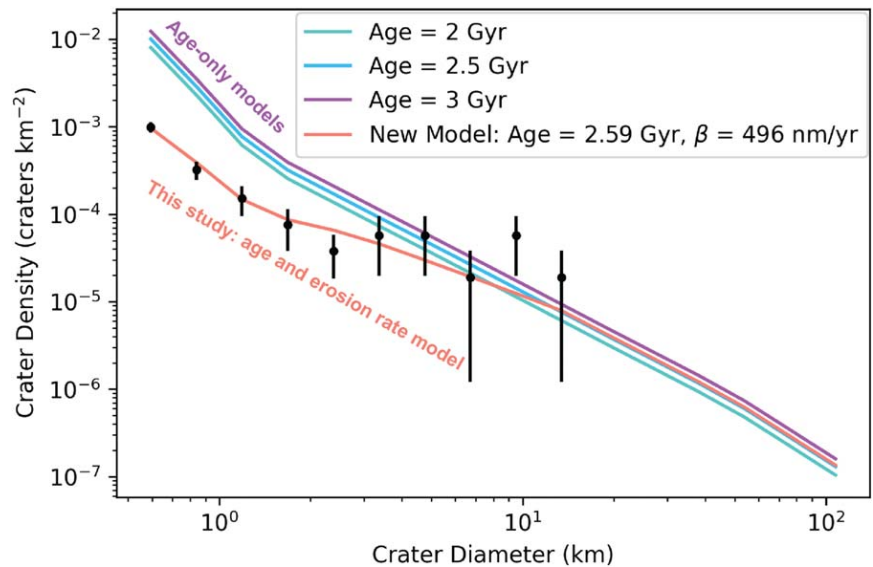

**Figure 2.** We apply a probabilistic model to find the best-fit age and $\beta$ for young sedimentary rock. Most previous models, with the exception of that of Palucis et al. (2020), do not account for both age and erosion rate. The incremental crater density distribution (black points) corresponds to site *4b* (in Figure 4(e)). Our new model's best fit for age and erosion rate is plotted against the data. The $1\sigma$ error for age is 1.47–3.76 Gyr, and for erosion rate it is 350–580 nm yr$^{-1}$. On the left, all three lines with negligible rates of erosion ($\beta = 1$ nm yr$^{-1}$) overestimate the crater density for bins with small crater diameters. Including the erosion rate in our new two-parameter model allows for a better fit. See Figure 3 for a visualization of why smaller craters are eroded more quickly.

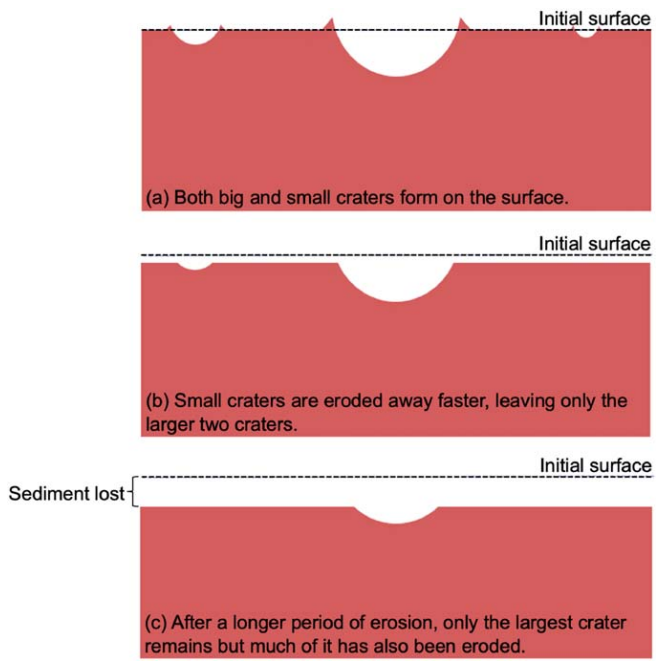

**Figure 3.** A simple diagram of the effect of erosion on crater removal.

In the next section (Section 2), we explain our new modeling technique and our new crater count data set. We then present the results marginalizing only age or erosion rate (Section 3.1), the results from the two-parameter model including both age and erosion rate (Section 3.2), and the best-fit predictions for each site compared with the data (Section 3.3). Then we discuss the overall trends for the 14 study sites (Sections 3.4, 4.1), the potential model uncertainties (Section 4.2), a $\chi^2$ test comparing the one-parameter and two-parameter models (Section 4.3), the implications for a shallow subsurface ice table near the equator of Mars (Section 4.4), and the implications for the global sedimentary cycle on Mars (Section 4.5). We conclude in Section 5.

## 2. Methods

### 2.1. Crater Count Data

We focused on 14 sites within the MFF and other nearby sites that are believed to be young sedimentary rock (Figure 4). The craters for each site were counted by one of us using Mars Reconnaissance Orbiter Context Camera (CTX) imagery. Figure 5 provides an example of two counted craters. Obvious secondary craters were excluded during crater counting, such as highly elliptical craters, very clustered craters, and secondary craters concentrated in rays coming from primary craters (Robbins & Hynek 2014).

We initially defined our site boundaries based on geological units of Mars (Tanaka et al. 2014) and, where available, further distinguished units into subcomponents, such as "a," "b," and "c," by using a combination of radar reflections, age estimates from previous works (e.g., Kite et al. 2015), and clear visual differences in crater densities. For Central Medusae Fossae 2 and the Far East Medusae Fossae, we used only the global geologic map (Tanaka et al. 2014) to separate distinct geologic units. For Central Medusae Fossae 1, sites *1a–1c* (Figure 4(b)), known as Lucus Planum, our sites *1a–1c* correspond to the Amazonian and Hesperian transition undivided unit (AHtu) in Tanaka et al. (2014). We do not include the area to the left of sites *1a–1c* (which is the focus of Mittelholz et al. 2020 ) because it corresponds to the older Hesperian transition undivided unit (Htu; Tanaka et al. 2014). Orosei et al. (2017) identified distinct areas within Lucus Planum based on radargrams of the subsurface. Since these areas with distinct radargrams also match variations in surface morphology (Orosei et al. 2017), we decided to assess these areas independently from each other for the age and erosion rate. Central Medusae Fossae 1 C (site *1c*) has similar boundaries to a large lobe northeast of Memnonia Sulci, labeled "C" in Figure 4 of Orosei et al. (2017). Similarly, we defined Central Medusae Fossae 2 B (site *4b*) based on lobe "B" in Figure 4 of Orosei et al. (2017). By taking radargrams of Lucus Planum, Orosei et al. (2017) found "B" and "C" as areas with concentrated radar reflectors, unlike the central section of Lucus Planum. As a result, both "B" and "C" indicate differences within the subsurface material at Lucus Planum relative to areas of no radar signal (Orosei et al. 2017). For Zephyria Planum, since most of Aeolis Dorsa has a relatively old age (Figure S1 in Kite et al. 2014; Zimbelman & Scheidt 2012; Burr et al. 2021), we restricted our count to the relatively young Zephyria Planum (unit "Y" in Kite et al. 2015). Sites *6a* and *6b*, East MFF, were separated based on visual inspection of the crater density.

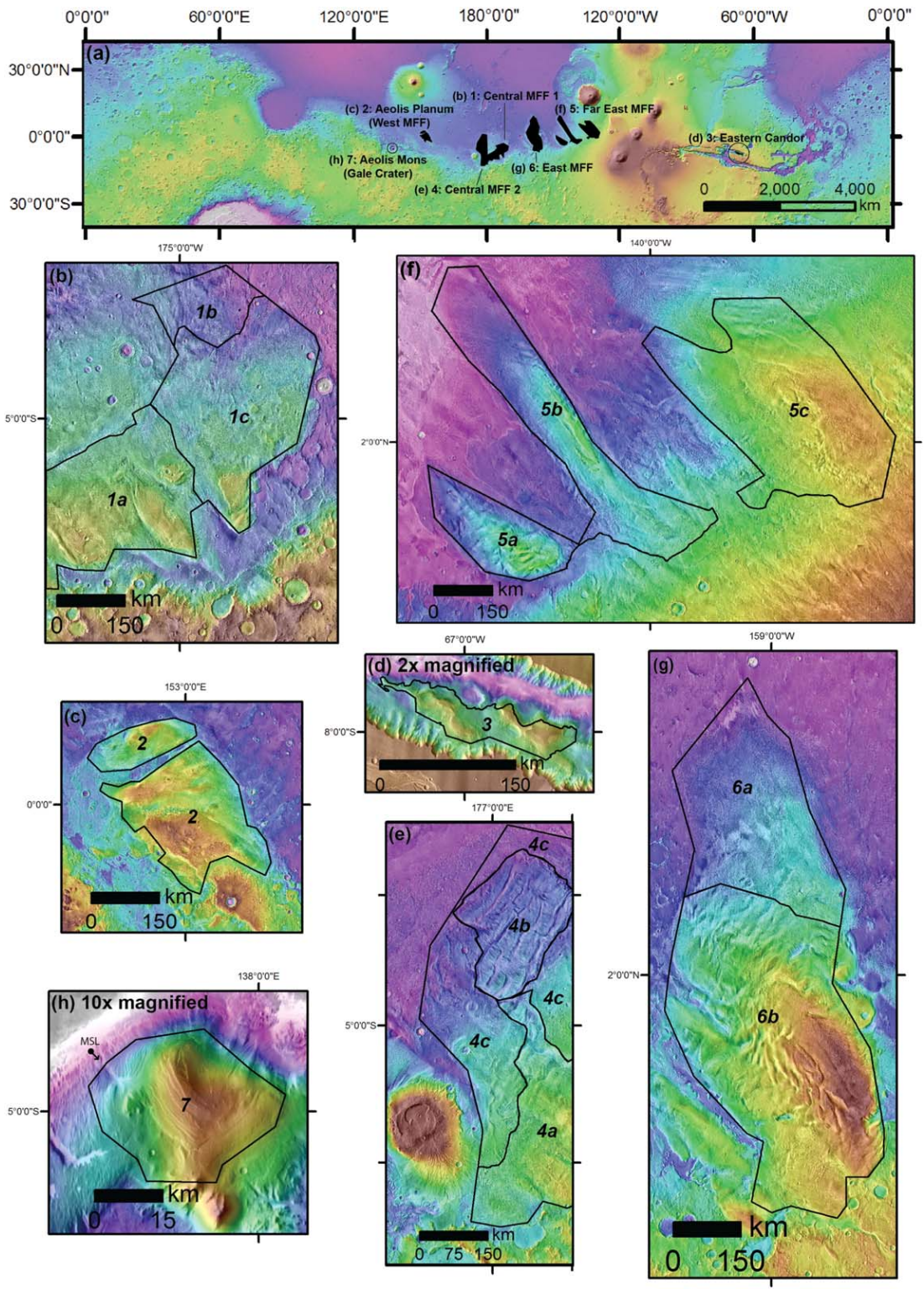

**Figure 4.** (a) The locations of the 14 sites, including the MFF, Upper Mount Sharp, and Eastern Candor. The colored topography map is from the Mars Orbiter Laser Altimeter (MOLA; Smith et al. 2001). Circles are placed around sites *3*: Eastern Candor and *7*: Aeolis Mons for visibility. For location (a), the lowest MOLA elevation is −8 km and the highest is 22 km. For locations (b)–(h), the base map is from the Thermal Emission Imaging System daytime IR 100 meter/pixel mosaic (Edwards et al. 2011). (b) Central MFF 1 includes the sites *1a*, *1b*, and *1c*. The MOLA elevations range from −4.3 km to 1.2 km. (c) Zephyria Planum (West MFF) is site 2. The MOLA elevations range from −3.3 km to −1.2 km. (d) Eastern Candor is site *3*. The MOLA elevations range from −5.4 km to 5.7 km. (e) Central MFF 2 includes the sites *4a*, *4b*, and *4c*. The MOLA elevations range from −3.9 to 3.2 km. (f) The Far East MFF includes the sites *5a*, *5b*, and *5c*. The MOLA elevations range from −3.9 km to 3.6 km. (g) The East MFF includes the sites *6a* and *6b*. The MOLA elevations range from −4.5 km to 0.6 km. (h) Upper Mount Sharp is site *7*. The MOLA elevations range from −4.5 km to 0.5 km. The black point labeled "MSL" corresponds to where the Mars Science Laboratory Curiosity rover is approaching Upper Mount Sharp.

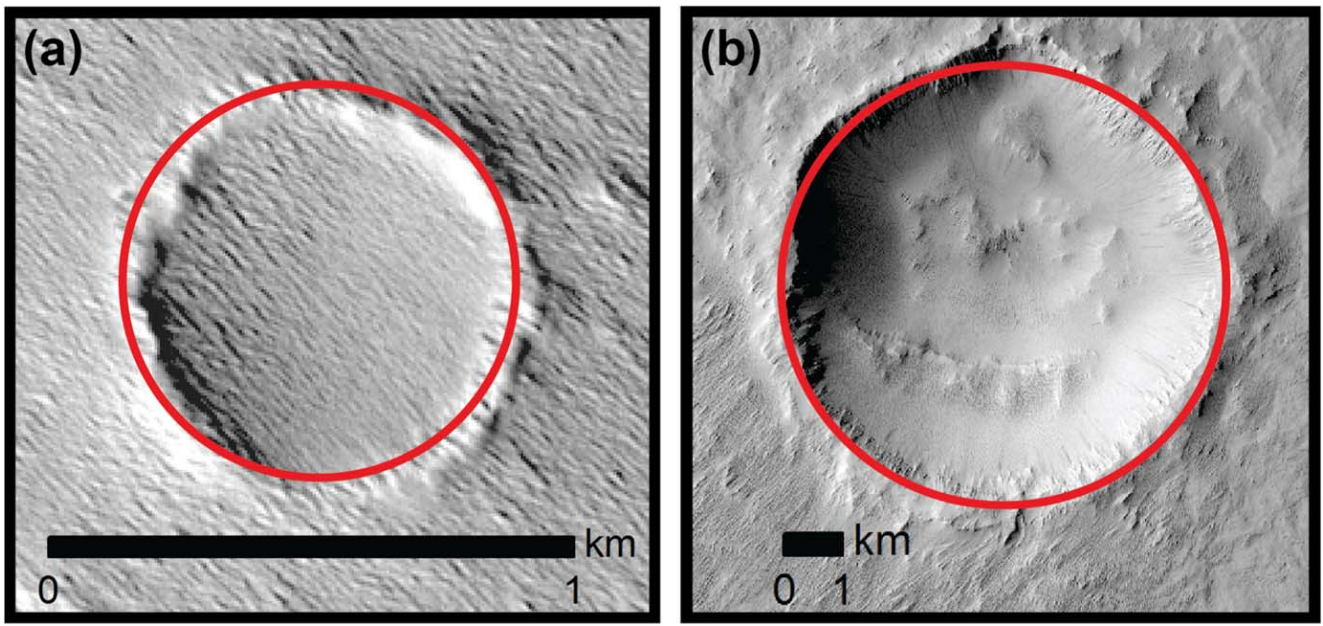

**Figure 5.** The diameters of the red circles correspond to the diameters of the craters. The crater images are from CTX data (Malin et al. 2007). (a) A crater from site *5*, Far East Medusae Fossae A (−136.6°E, −0.7°N). (b) A crater from site *1*, Central Medusae Fossae 1 A (−174.8°E, −2.1°N).

By plotting the logarithm of the number of counted craters in the data set for each given minimum diameter, we found the beginning of the diameter rolloff to be 0.510 km. This rollover at approximately 0.5 km represents where the crater counts are affected by survey incompleteness. As a result, we set 0.5 km as the minimum bin diameter for the smallest crater bin. We use manually counted craters within each of the 14 sites and divide the craters into bins, based on crater diameters 0.5–128 km, for a total of 16 bins that are spaced by a factor of about $2^{-1/2}$ between each consecutive minimum diameter (Michael 2013). The Hartmann isochrons in Table 1 of Michael (2013) are defined as the size distribution of craters for a surface with a specified age and without erosion processes, based on a widely used Mars crater chronology model (Hartmann 2005).

### 2.2. Model Description

Previous models typically only consider crater count data in order to estimate the absolute age of the sedimentary areas. The next step is to create a two-parameter model that addresses both erosion rate and age, which is the route that we explore in this study. The erosion rate is set as a fixed rate (after a single episode of deposition) without interruption because this is the simplest treatment of erosion. The reality is likely more complicated (Section 1). Our data set includes a total of 2119 craters, too few to merit a more complicated model. However, future work incorporating larger areas and populations of craters could investigate using an erosion rate that varies with time. We assume that there was a single episode of sedimentation, corresponding to the age of the mound when it was initially deposited. As discussed in Section 2.1, our crater data set is limited to 0.5 km, due to incompleteness. However, this larger diameter allows this work to focus on the bulk age of the MFF, since smaller craters would only be sensitive to the uppermost MFF. In our model, we assume that craters are generated randomly as a Poisson process, given by the equation

$$P(n) = \frac{e^{-\mu}\mu^n}{n!}, \qquad (1)$$

where $P(n)$ is the Poisson probability of observing $n$ craters and $n$ is the observed number of craters, as determined by the crater counts. Our model fits an expected number of craters, $\mu$, and estimates $\mu$ without erosion and with erosion. The model then chooses the smaller $\mu$ between the no-erosion and erosion cases. The expected number of craters for no erosion is

$$\mu = Hat, \qquad (2)$$

where $H$ is the expected number of craters for each crater diameter bin (craters km$^{-2}$ Gyr$^{-1}$), taken from Table 1 in Michael (2013), which is based on the Hartmann production function (Hartmann 2005). $a$ is the area of the crater count site (km$^2$) and $t$ is the age (Gyr).

The expected number of craters with erosion is:

$$\mu = Had/\beta. \qquad (3)$$

$H$ is still the expected number of craters for each crater diameter bin (craters km$^{-2}$ Gyr$^{-1}$) and $a$ is the area of the crater count site (km$^2$), as mentioned above for the no-erosion case. $\beta$ is the inputted trial erosion rate (nm yr$^{-1}$). $d$ is the logarithmic mid-depth, based on the minimum and maximum diameter for each specific crater bin. The depth is related to diameter as follows:

$$d = \begin{cases} 0.2D & \text{if } D < 2.82 \text{ km} \\ 0.323D^{0.538} & \text{if } D \geqslant 2.82 \text{ km} \end{cases}. \quad (4)$$

$D$ is the logarithmic mid-diameter of the crater bin in meters:

$$D = \sqrt{D_{\min}D_{\max}}. \quad (5)$$

We adapted pre-existing depth-to-diameter equations to better fit the observed depth-to-diameter relationships (such as the idealized cases mentioned in Gabasova & Kite 2018 and Watters et al. 2015). For small craters ($D < 2.82$ km), we used $d = 0.2$D, which was first determined by Pike (1980) and reconfirmed by Watters et al. (2015). This depth-to-diameter relationship was also used by both Palucis et al. (2020) and Smith et al. (2008). For large craters ($D \geqslant 2.82$ km), the component $0.323D^{0.538}$ in Equation (4) is from the Mars data set of Tornabene et al. (2018). Tornabene et al. (2018) use this depth–diameter scaling relationship for diameters larger than 12 km. We chose 2.82 km as the transition diameter because it is the intersection between the two equations $d = 0.2D$ and $d = 0.323D^{0.538}$. As a result, we adjusted the cutoff to 2.82 km, to make sure the depth estimates were continuous between the two equations and monotonically increasing. In previous work, Pike (1980) used a transition of $D = 1$ km, while Palucis et al. (2020) and Smith et al. (2008) used the 0.2$D$ relationship with a transition at $D = 5.8$ km.

The $\mu$ for both the erosion and no-erosion cases are also multiplied by a correction factor to account for the increased impact flux early in Martian history. Based on Table 1 from Michael (2013), $N$ is the expected number of craters in the last column (craters km$^{-2}$ Gyr$^{-1}$). Taking the cumulative density $N$ at 1 km from Table 1, we have $N_{steady} = 5.84 \times 10^{-4}t$, which incorrectly assumes that the modern impact flux over the past 10 million yr has always held. Empirical estimates of the modern impact flux are based on satellite observations of newly appeared craters (Daubar et al. 2013), and vary from model predictions by about a factor of 4 (Daubar et al. 2014). This observed present-day crater flux is uncertain due to the spatial nonrandomness of the observations (Daubar et al. 2013), since the true flux is expected to be almost spatially random (Le Feuvre & Wieczorek 2008; Kite & Mayer 2017). In addition, satellite observations may not reflect the recent changes in flux throughout the different orbital cycles of Mars over the past 10 million yr (Kite & Mayer 2017). As a result, we focus on the Michael (2013) flux, instead of a flux derived from empirical observations.

Adapting Equation (3) from Michael (2013) states that $N = 3.79 \times 10^{-14}(e^{6.93\,t} - 1) + 5.84 \times 10^{-4}t$. This gives the correction factor

$$\frac{N}{N_{\text{steady}}} = \frac{3.79 \times 10^{-14}(e^{6.93t} - 1) + 5.84 \times 10^{-4}t}{5.84 \times 10^{-4}t} \quad (6)$$

for the case without erosion. For the case with erosion, $t$ is replaced by $d/\beta$, where $d$ is again the logarithmic mid-depth (m) for a given crater bin and $\beta$ is again the erosion rate (nm yr$^{-1}$). The uncertainties in this correction factor are discussed in Section 4.2.

For our model, the range of ages considered were 1 s to 4.5 Gyr, with 1000 ages linearly spaced in between. Similarly, $\beta$ ranged from 1 nm yr$^{-1}$ to 2000 nm yr$^{-1}$, with 1000 $\beta$s linearly spaced in between (higher values are considered for the fastest-eroding sites). Taking each array of $P(n)$ values from the individual crater diameter bins, we find the total probability as the product of all 16 bins:

$$P_{\text{total}}(t, \beta) = \prod_{n=1}^{16} P(n). \quad (7)$$

In addition to evaluating the two-parameter probability distribution, we also collapsed the two-parameter probability array, by summing along one dimension (either $t$ or age) to find the normalized one-parameter probability array for age and for $\beta$. Then, for the one-parameter cumulative probability array, we found the cumulative sum of the probability array and normalized it to a maximum cumulative probability of one. The median of the cumulative distribution (for either age or $\beta$) corresponds to the best fit for the one-parameter model.

From the two-parameter model, we calculated the crater density predicted by the best fit, which corresponds to the age and $\beta$ pair that has the highest predicted probability from our model. Using the best-fit age and $\beta$, we create a model-generated CSFD to compare with the real data's CSFD. For the model-generated CSFD, we found the expected crater density per km$^2$ for each crater diameter bin by using Equations (2) and (3). The error bars for the crater densities came from the 1$\sigma$ standard deviation from the center of the cumulative distribution, in which n, the observed number of craters, is equal to $\mu$, the expected number of craters for the crater count data. The model Python scripts for generating the figures are available via the link in the Appendix.

### 2.3. Model Tests

To assess whether our model is accurate and to examine the uncertainty of our model's predictions, we generate synthetic crater distributions based on Table 1 of Michael (2013) for a given age and erosion rate. We then apply our model to estimate the best fit for age and erosion rate, and compare our model's result with the known inputs.

For test 1 (Figure 6(a)), we first found the expected number of craters per bin using Table 1 of Michael (2013; based on Hartmann 2005) and Equations (2) and (3). The input parameters for test 1 were age = 4 Gyr, erosion rate = 1 nm yr$^{-1}$, and area = 10$^5$ km$^2$. The expected craters per bin (corresponding to the Table 1 bin indices −2–13 in Michael 2013) rounded to the nearest whole number were [30124, 8618, 2308, 959, 513, 275, 148, 79, 43, 23, 12, 7, 3, 2, 1, and 0]. From Figure 6(a), the 1$\sigma$ range of ages was narrow between 3.99 and 4.01 Gyr, whereas the 1$\sigma$, 2$\sigma$, and 3$\sigma$ uncertainty regions for $\beta$ almost reached 30 nm yr$^{-1}$, larger by almost a factor of 30 than the input $\beta = 1$ nm yr$^{-1}$. This is because small but non-negligible erosion rates cannot be ruled out without data for $D < 500$ m craters, which were not

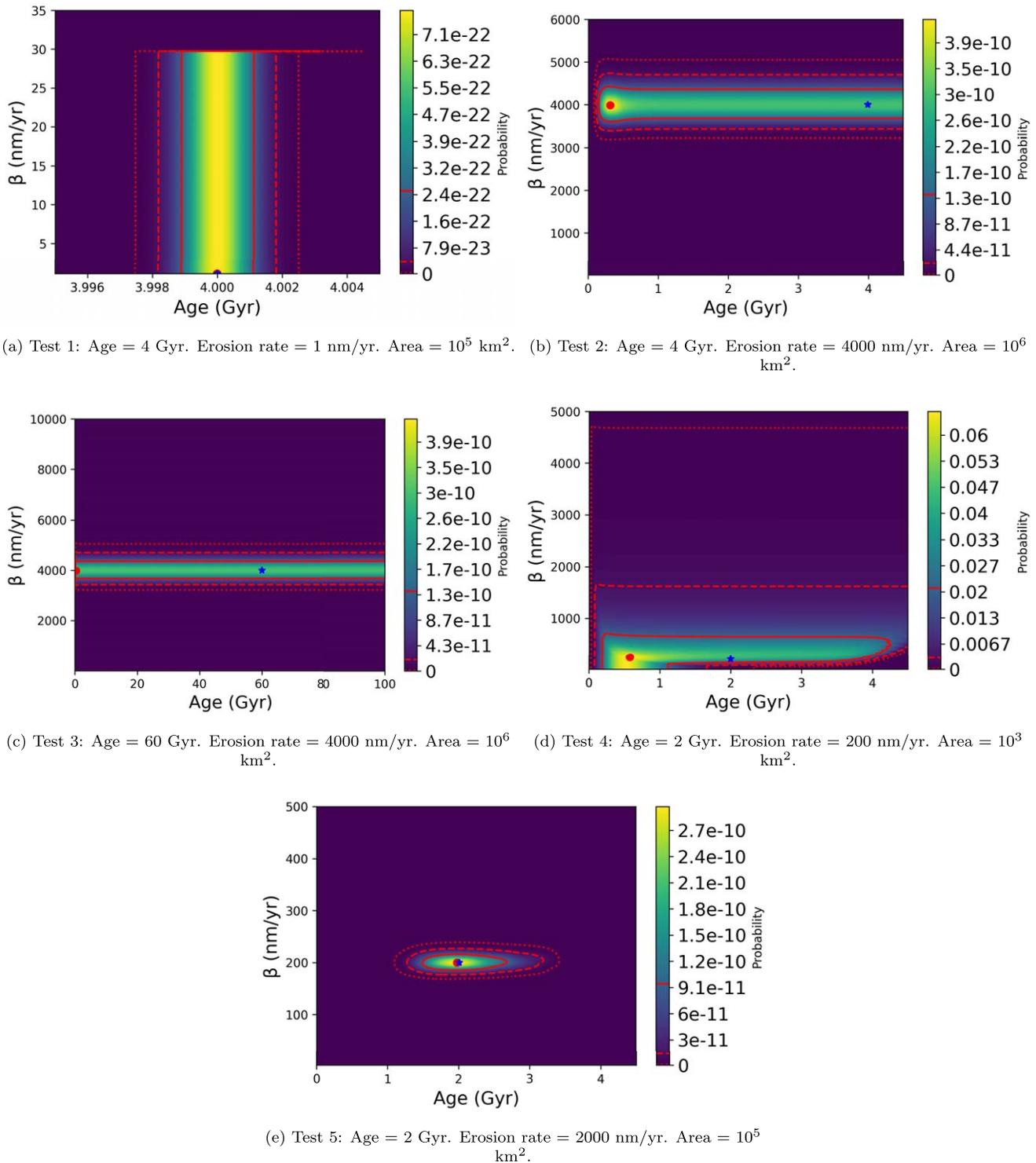

(a) Test 1: Age = 4 Gyr. Erosion rate = 1 nm/yr. Area = $10^5$ km$^2$.

(b) Test 2: Age = 4 Gyr. Erosion rate = 4000 nm/yr. Area = $10^6$ km$^2$.

(c) Test 3: Age = 60 Gyr. Erosion rate = 4000 nm/yr. Area = $10^6$ km$^2$.

(d) Test 4: Age = 2 Gyr. Erosion rate = 200 nm/yr. Area = $10^3$ km$^2$.

(e) Test 5: Age = 2 Gyr. Erosion rate = 2000 nm/yr. Area = $10^5$ km$^2$.

**Figure 6.** Five injection tests for the model, with synthetic data generated from different age, erosion rate, and area inputs. The blue star corresponds to the input age and erosion rate, while the red circle corresponds to the two-parameter model's best-fit age and erosion rate, based on the highest probability. The solid line is the $1\sigma$ uncertainty on the retrieved parameters, the dashed line is $2\sigma$, and the dotted line is $3\sigma$. The differences in magnitude between the probability values of different sites should not be compared across plots; rather, the probability values should only be considered relative to each individual site.

included in the test. In Figure 6(a), the yellow region of highest probability contained the blue star, corresponding to the input parameters, and the red circle, the model's best-fit age and erosion rate, accurately predicted the input parameters at 4 Gyr and 1 nm yr$^{-1}$.

The input parameters for test 2 (Figure 6(b)) were age = 4 Gyr, erosion rate = 4000 nm yr$^{-1}$, and area = $10^6$ km$^2$. The expected craters per bin were [120, 48, 18, 11, 8, 6, 4, 2, 2, 1, 1, 0, 0, 0, and 0]. From Figure 6(b), we see that age, as expected, was not well constrained. The yellow region of highest probability, with the red circle for the best-fit age of 0.31 Gyr and erosion rate of 3990 nm yr$^{-1}$, was offset from the blue star representing the input parameters. Nevertheless, the $1\sigma$ region, which enclosed 68% of the total probability, did contain the blue star.

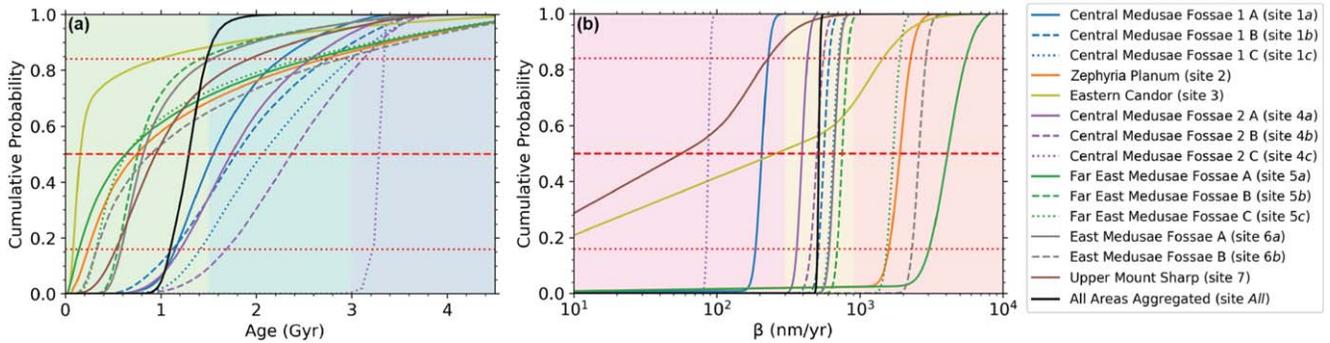

**Figure 7.** The dashed red line corresponds to the median of the cumulative distribution and the red dotted lines correspond to the $1\sigma$ error range. (a) Age CDFs for all 14 sites. The green, teal, and dark blue background colors correspond to the age increments used in Figure 10. (b) $\beta$ CDFs for all 14 sites. The pink, yellow, and orange-red background colors correspond to the $\beta$ increments used in Figure 10.

The input parameters for test 3 (Figure 6(c)) were age = 60 Gyr, erosion rate = 4000 nm yr$^{-1}$, and area = $10^6$ km$^2$, corresponding to an unrealistically old surface, and assuming the same relationship for impact flux as described in Section 2.2. The expected craters per bin were [120, 48, 18, 11, 8, 6, 4, 2, 2, 1, 1, 0, 0, 0, 0, and 0]. Holding the erosion rate and area constant, but changing the age in comparison with test 2, the expected craters per bin were still the same as those in test 2. This is because the high erosion rate of 4000 nm yr$^{-1}$ prevents the older surface in test 3 from having more craters than test 2. From Figure 6(c), we see that age was not constrained at all, as expected. While the best-fit erosion rate was 3980 nm yr$^{-1}$, and corresponded well to the input erosion rate of 4000 nm yr$^{-1}$, the best-fit age was only 0.30 Gyr. However, the $1\sigma$ uncertainty region contained the blue star, corresponding to the input parameters.

The input parameters for test 4 (Figure 6(d)) were age = 2 Gyr, erosion rate = 200 nm yr$^{-1}$, and area = $10^3$ km$^2$, corresponding to a small area, which is the approximate area of site 7, Upper Mount Sharp. The expected craters per bin (corresponding to Michael 2013 Table 1 bin indices −2 to 13) were [2, 1, 0, 0, 0, 0, 0, 0, 0, 0, 0, 0, 0, 0, 0, 0]. From Figure 6(d), we see that age was not well constrained, ranging from 0 to 4.5 Gyr for the $3\sigma$ region. This is likely due to the minimal number of craters. Maximum $\beta$ for $1\sigma$ versus $3\sigma$ varied greatly, from 710 nm yr$^{-1}$ to 4730 nm yr$^{-1}$. The yellow region of highest probability with a best-fit age of 0.58 Gyr did not include the input parameters, but the $1\sigma$ region did. In addition, the best-fit erosion rate of 240 nm yr$^{-1}$ did approximate the input erosion rate of 200 nm yr$^{-1}$.

The input parameters for test 5 (Figure 6(e)) were age = 2 Gyr, erosion rate = 200 nm yr$^{-1}$, and area = $10^5$ km$^2$, corresponding to the approximate area of Central Medusae Fossae 1 and the age and erosion rate predicted for the site (Figure 6(e)). The expected craters per bin (corresponding to the Table 1 bin indices −2 to 13 in Michael 2013) were [239, 97, 37, 21, 14, 7, 4, 2, 1, 1, 0, 0, 0, 0, 0, and 0]. The $1\sigma$ region had an age above 1.48 Gyr and below 2.67 Gyr, with an erosion rate between 190 and 220 nm yr$^{-1}$. The $3\sigma$ region estimated an age between 1.08 and 3.43 Gyr, with a similar erosion rate to the $1\sigma$ region between 170 and 240 nm yr$^{-1}$. The region of highest probability contained the blue star, and the best-fit age of 1.96 Gyr was very close to the input age of 2 Gyr and the best-fit erosion rate was the same as the input.

The tests reveal a few reasons as to why age might be unconstrained. For example, tests 2 and 3, which had the same inputs for area and erosion rate, but different ages, resulted in the same amount of expected craters per bin and, correspondingly, wide error ranges for age. This reveals that in cases of high erosion rates, such as 4000 nm yr$^{-1}$ in tests 2 and 3, it is difficult for the model to tightly constrain the age. In addition, test 4 is an example of a small area with few craters. Due to the small amount of craters, it is a challenge for the model to pinpoint age, and the size of the error regions of the erosion rate also increases. As a result, the tests show that the erosion rate is more tightly constrained than age in cases of small areas and/or high erosion rates. Nevertheless, since the input parameters for both age and erosion rate (the blue star) in Figure 6 were always contained within the $\sigma$ error regions, these tests demonstrate that the model is reliable in constraining age and erosion rate within these error regions.

## 3. Results

### 3.1. One-parameter Posterior Probability Distribution Functions

After collapsing the two-parameter probability contour plot along $\beta$ to form a one-parameter cumulative distribution function (CDF) of age in Figure 7(a), we compared the age probability distributions for each region. The best-fit absolute model ages for 13 of the 14 regions fall below 2.5 Gyr, and 11 of the 14 sites have $1\sigma$ ages that are younger than 3 Gyr. Four of the sites have $1\sigma$ error regions that are completely constrained below 2.5 Gyr. Eastern Candor (3), in particular, is predicted to be young because its $1\sigma$ error region is constrained below 1 Gyr. All six Central Medusae Fossae sites (1a–1c and 4a–4c) have lower $1\sigma$ error bounds that are older than 1 Gyr and upper $1\sigma$ error bounds that are between 2.0 and 3.2 Gyr, which makes the Central Medusae Fossae sites the oldest among the sedimentary mounds examined in this study.

In Figure 7(b), well-constrained cumulative probability curves for $\beta$ (with tight $1\sigma$ error bounds) include all sites, except Eastern Candor (3) and Upper Mount Sharp (7). Most sites (excluding Eastern Candor and Upper Mount Sharp) require large erosion rates, at least greater than 10 nm yr$^{-1}$. It is likely that Eastern Candor and Upper Mount Sharp also require erosion rates >10 nm yr⁻¹ because although their lower 1σ are not bound above 10 nm yr⁻¹, both the median and upper 1σ are bound above 10 nm yr⁻¹. As a result, with 1σ certainty, we can say that 12 of the 14 sites have non-negligible erosion rates.

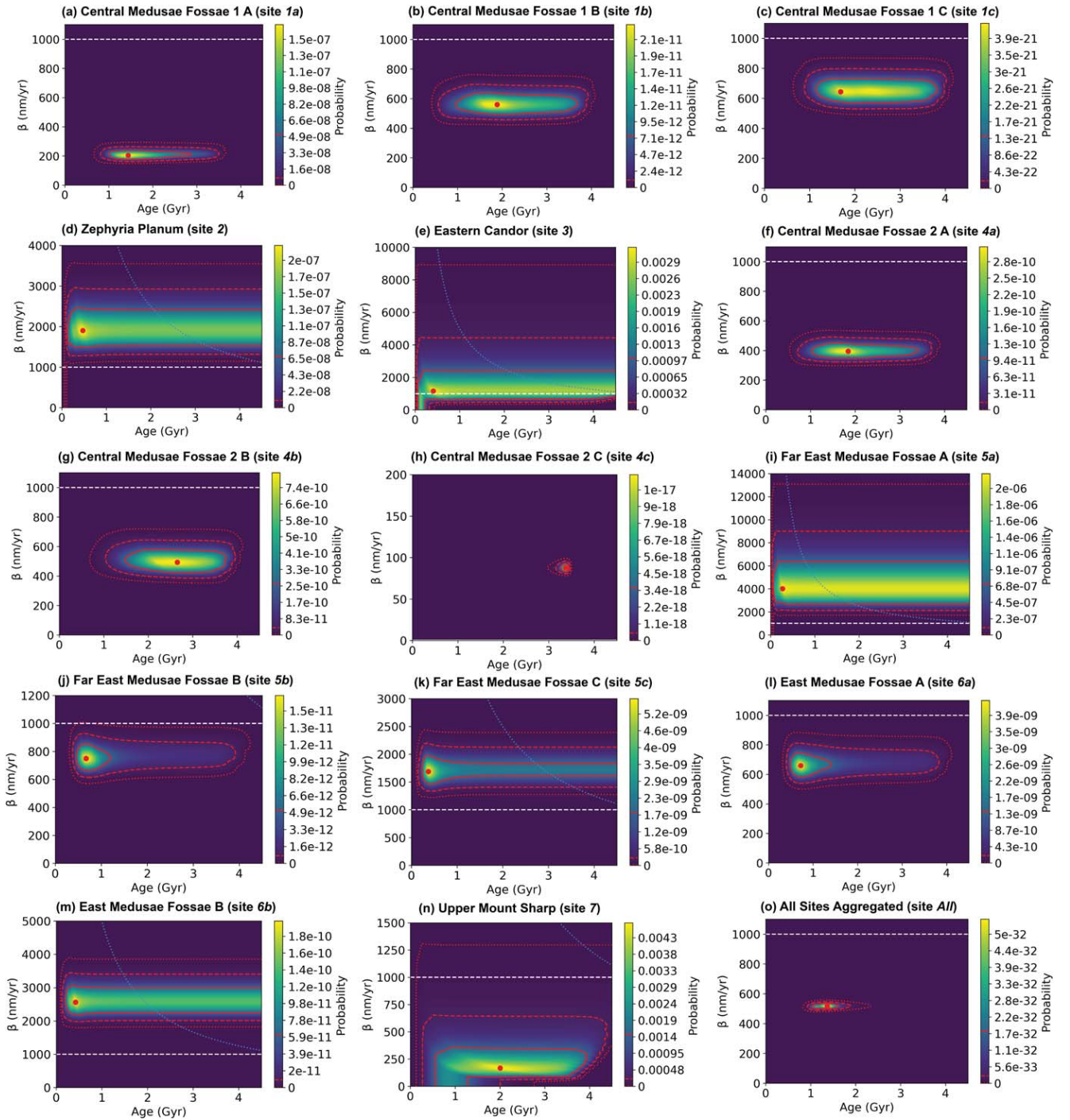

**Figure 8.** For the plots of each site, the red circle corresponds to the two-parameter model's best-fit age and erosion rate, based on the highest probability; the solid red line represents the 1σ uncertainty region; the dashed red line represents the 2σ uncertainty region; and the dotted red line represents the 3σ uncertainty region. The yellow regions highlight the area of most probable age and β, while the darkest purple regions indicate the least likely age and β. The y-axis scale for β varies between each area, to best showcase the probability region. The dashed white line at 1000 nm yr⁻¹ indicates the minimum erosion rate needed for metastable water ice at depths shallow enough to be detected by GRS or NS. The value of 1000 nm yr⁻¹ is based on the right dashed line of Figure 2 from Hudson et al. (2007), as explained in Section 4.4, and represents the minimum erosion rate necessary for a shallow subsurface water ice table. This means that if β is greater than 1000 nm yr⁻¹, the depth of the buried ice will be less than 1 m. The dotted blue line is a reference value for 5 km of erosion at each given age and β value. The differences in magnitude between the probability values of the different sites should not be compared across plots; rather, the probability values should only be considered relative to each individual site.

### 3.2. Two-parameter Model Output

All two-parameter contour plots are presented in Figure 8 and the model predictions are summarized in Table 2. The site-specific results are discussed below.

**Table 2**
Summary of the Two-parameter Model Predictions for Age and $\beta$, Including Best-fit Estimates and $1\sigma$, $2\sigma$, and $3\sigma$ Uncertainty Ranges

| Name | Site | Best Fit | | $1\sigma$ | | $2\sigma$ | | $3\sigma$ | |
|---|---|---|---|---|---|---|---|---|---|
| | | t (Gyr) | $\beta$ (nm yr$^{-1}$) | t (Gyr) | $\beta$ (nm yr$^{-1}$) | t (Gyr) | $\beta$ (nm yr$^{-1}$) | t (Gyr) | $\beta$ (nm yr$^{-1}$) |
| *Central Medusae Fossae 1* | *1* | *2.19* | *490* | *1.43–3.39* | *450–540* | *1.12–3.64* | *430–570* | *0.84–3.80* | *410–600* |
| Central Medusae Fossae 1 A | *1a* | 1.43 | 210 | 1.00–2.88 | 180–240 | 0.81–3.51 | 160–270 | 0.66–3.66 | 150–290 |
| Central Medusae Fossae 1 B | *1b* | 1.89 | 560 | 0.95–3.68 | 490–640 | 0.57–3.94 | 460–700 | 0.42–4.09 | 420–760 |
| Central Medusae Fossae 1 C | *1c* | 1.67 | 650 | 1.13–3.82 | 570–730 | 0.88–4.03 | 530–800 | 0.64–4.18 | 490–870 |
| Zephyria Planum | *2* | 0.47 | 1910 | 0.12–4.50 | <2470 | 0.09–4.50 | <2970 | 0.06–4.50 | <3580 |
| Eastern Candor | *3* | 0.42 | 1140 | 0.07–4.50 | <2470 | 0.04–4.50 | <4490 | 0.02–4.50 | <8970 |
| *Central Medusae Fossae 2* | *4* | *3.18* | *160* | *2.54–3.36* | *140–170* | *2.28–3.42* | *140–170* | *2.04–3.47* | *140–180* |
| Central Medusae Fossae 2 A | *4a* | 1.81 | 390 | 1.02–3.42 | 350–450 | 0.71–3.72 | 320–490 | 0.55–3.90 | 300–540 |
| Central Medusae Fossae 2 B | *4b* | 2.59 | 500 | 1.47–3.76 | 420–580 | 1.04–3.97 | 390–650 | 0.69–4.13 | 350–720 |
| Central Medusae Fossae 2 C | *4c* | 3.41 | 90 | 3.26–3.44 | 80–90 | 3.15–3.48 | 80–100 | 2.98–3.51 | 80–100 |
| *Far East Medusae Fossae* | *5* | *0.49* | *1260* | *0.33–4.50* | *1140–1380* | *0.27–4.50* | *<1080–1480* | *0.22–4.50* | *1020–1570* |
| Far East Medusae Fossae A | *5a* | 0.27 | 4010 | 0.02–4.50 | <6410 | 0.04–4.50 | <9040 | 0.06–4.50 | <13200 |
| Far East Medusae Fossae B | *5b* | 0.68 | 750 | 0.46–1.20 | 660–850 | 0.38–3.93 | 610–930 | 0.31–4.24 | 570–1020 |
| Far East Medusae Fossae C | *5c* | 0.37 | 1690 | 0.24–4.50 | 1450–1980 | 0.15–4.50 | 1330–2200 | 0.11–4.50 | 1220–2440 |
| *East Medusae Fossae* | *6* | *0.54* | *1320* | *0.31–4.50* | *1180–1450* | *0.24–4.50* | *1110–1560* | *0.21–4.50* | *1040–1670* |
| East Medusae Fossae A | *6a* | 0.73 | 660 | 0.49–1.41 | 580–750 | 0.40–3.83 | 540–810 | 0.32–4.12 | 500–890 |
| East Medusae Fossae B | *6b* | 0.42 | 2590 | 0.09–4.50 | 2180–3010 | 0.12–4.50 | 1970–3450 | 0.20–4.50 | 1790–3890 |
| Upper Mount Sharp | *7* | 2.04 | 170 | 0.55–3.95 | <360 | 0.27–4.40 | <660 | 0.15–4.50 | <1310 |
| All Sites Aggregated | *All* | 1.37 | 520 | 1.03–1.61 | 500–540 | 0.93–1.90 | 490–550 | 0.83–2.34 | 480–560 |

**Note.** Ages have been rounded to the nearest 10 Ma and $\beta$s have been rounded to the nearest 10 nm yr$^{-1}$. For the locations, see Figure 4. The results for sites *1*, *4*, *5*, and *6* have been italicized and are included for the purpose of comparison with their subregions in Section 4.2.

Central Medusae Fossae 1 (**A**: $t = 1.43^{+1.45}_{-0.43}$ Gyr, $\beta = 210 \pm 30$ nm/yr; **B**: $t = 1.89^{+1.79}_{-0.94}$ Gyr, $\beta = 560^{+80}_{-70}$ nm yr$^{-1}$; **C**: t = $1.67^{+2.15}_{-0.54}$ Gyr, $\beta = 650 \pm 80$ nm yr$^{-1}$) (Figures 8(a)–(c))

1. The estimated age ranges are similar between these three sites.
2. The $\beta$ values vary greatly: sites B and C have $\beta$ two or more times greater than that of site A.

Zephyria Planum ($t = 0.47^{+4.03}_{-0.35}$ Gyr, $\beta = 1910^{+560}_{-1910}$ nm yr$^{-1}$) (Figure 8(d))

1. While the $1\sigma$ region for $\beta$ was predicted to be between 0 and 2470 nm yr$^{-1}$, age was not constrained.
2. With only 36 craters (Table 1), the paucity of craters is likely one reason why age and $\beta$ were not well constrained.

Eastern Candor ($t = 0.42^{+4.08}_{-0.35}$ Gyr, $\beta = 1140^{+1330}_{-1140}$ nm yr$^{-1}$) (Figure 8(e))

1. Similar to Zephyria Planum, at ages younger than 0.5 Gyr, the predicted $\beta$ values drop down to include very low values. In fact, all three error regions drop down to low $\beta$ values.
2. In the entire Eastern Candor site studied, there were only five craters, with four craters in the smallest crater bin

diameter size. This paucity of craters likely resulted in no useful age constraint.

Central Medusae Fossae 2 (**A**: $t = 1.81^{+1.61}_{-0.79}$ Gyr, $\beta = 390^{+60}_{-40}$ nm yr$^{-1}$; **B**: $t = 2.59^{+1.17}_{-1.12}$ Gyr, $\beta = 500^{+80}_{-80}$ nm yr$^{-1}$; **C**: $t = 3.41^{+0.03}_{-0.15}$ Gyr, $\beta = 90^{+0}_{-10}$ nm yr$^{-1}$) (Figures 8(f)–(h))

1. Site C of Central Medusae Fossae 2 had the most tightly constrained age and $\beta$ values of all the sites, with all three of the error ranges falling between 2.98 and 3.51 Gyr for age, and a best-fit prediction for $\beta$ between 80 and 100 nm yr$^{-1}$.

Far East Medusae Fossae (**A**: $t = 0.27^{+4.23}_{-0.25}$ Gyr, $\beta = 4010^{+2400}_{-4010}$ nm yr$^{-1}$; **B**: $t = 0.68^{+0.52}_{-0.22}$ Gyr, $\beta = 750^{+100}_{-90}$ nm yr$^{-1}$; **C**: $t = 0.37^{+4.13}_{-0.13}$ Gyr, $\beta = 1690^{+290}_{-240}$ nm yr$^{-1}$) (Figures 8(i)–(k))

1. While the best-fit age is young, we see that the $2\sigma$ uncertainty range is very wide, spanning a range of almost 4 billion yr.
2. Site C of Far East Medusae Fossae is similar to site B with a younger age estimated within the $1\sigma$ region, but a right tail permitting higher ages for larger-$\sigma$ uncertainty regions.

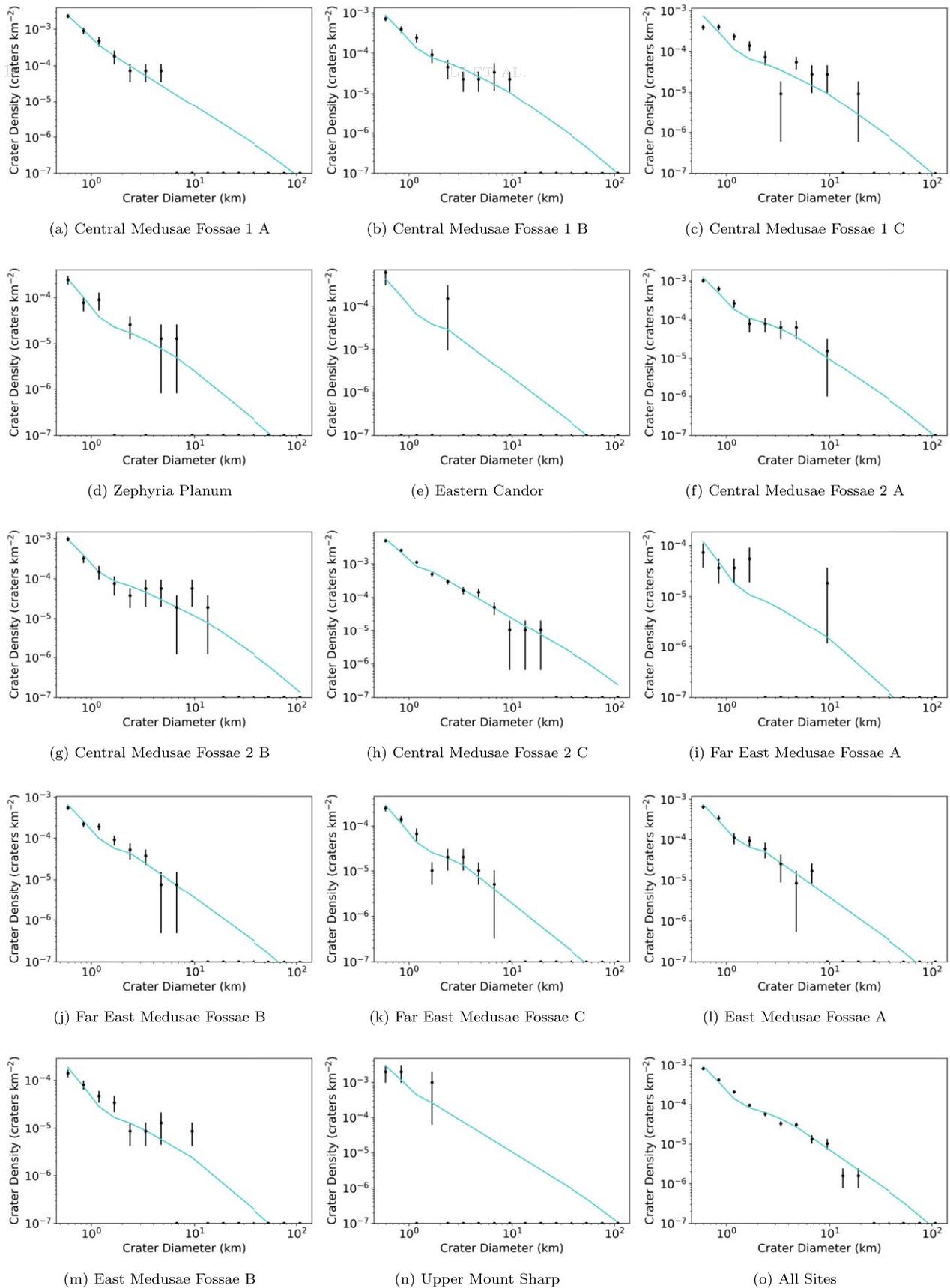

**Figure 9.** The cyan line in each plot represents the model-generated CSFD using the best-fit age and $\beta$ from the two-parameter model for each site. We plot this against the real CSFD, based on observed craters for 16 crater diameter bins. The observed data correspond to the black points. The black points marked on the $x$-axis with crater density values of $10^{-7}$ craters/km$^2$ correspond to the bins without craters. The black vertical bars represent the $1\sigma$ error range, by extrapolating from a cumulative distribution of each crater bin. The error bars of some bins are smaller than the corresponding disk symbols.

3. Site A has only 12 craters, so age and $\beta$ were not well constrained (Table 1).

East Medusae Fossae (**A**: $t = 0.73^{+0.68}_{-0.24}$ Gyr, $\beta = 660^{+90}_{-80}$ nm yr$^{-1}$; **B**: $t = 0.42^{+4.08}_{-0.33}$ Gyr, $\beta = 2590^{+420}_{-410}$ nm yr$^{-1}$) (Figures 8(l)–(m))

1. Both the Far East Medusae Fossae and East Medusae Fossae sites have younger best-fit ages, with older ages permitted at $2\sigma$.
2. Despite being next to each other, site A and site B have very different predicted erosion rates (580–750 nm yr$^{-1}$ versus 2180–3010 nm yr$^{-1}$).

Upper Mount Sharp ($t = 2.04^{+1.91}_{-1.49}$ Gyr, $\beta = 170^{+190}_{-170}$ nm yr$^{-1}$) (Figure 8(n))

1. The erosion rate estimate is broad. We also see that on the left-hand side of the plot, the uncertainty regions drop to include 1 nm yr$^{-1}$ for the erosion rate, indicating that the lowest erosion rate values only occur for the youngest ages. However, unlike previous sites, with the lowest $\beta$ values the youngest ages permitted here reach values as high as 2 Gyr ($2\sigma$).
2. In the entire Upper Mount Sharp site studied, there were only five craters (Table 1), with two craters in the two smallest crater bin diameter sizes. The paucity of craters accounts for the wide age range.

All 14 sites aggregated ($t = 1.37^{+0.24}_{-0.34}$ Gyr, $\beta = 520 \pm 20$ nm/yr) (Figure 8(o))

1. When combining the crater count data for all 14 sites, age and especially $\beta$ are tightly constrained within the framework of the model assumptions: ignoring regional variation (which is not a safe assumption), young sedimentary rocks on Mars are $1.4^{+0.2}_{-0.3}$ Gyr, with an erosion rate of $520 \pm 20$ nm yr$^{-1}$.

### 3.3. Best-fit Assessment

Figure 9 compares the observed crater distributions with the model's best-fit age and erosion rate for each site. From Figures 9(a) through 9(n), the sites with best-fit model predictions that were the most consistent with the crater data included Central Medusae Fossae 1 A (*1a*) and B (*1b*), Central Medusae Fossae 2 A (*4a*) and C (*4c*), Far East Medusae Fossae B (*5b*) and C (*5c*), East Medusae Fossae A (*6a*), and the aggregation of 14 sites. It is unclear whether the model was a good fit for sites such as Eastern Candor (*3*) in Figure 9(e), where there were relatively few filled crater bins to compare with the model.

### 3.4. Overall Trends for Each Site

#### 3.4.1. Central Medusae Fossae

A common trend throughout the two-parameter probability figures, one-parameter probability figures, and best-fit figures is the tight precision on Central Medusae Fossae 2 C (*4c*) in contrast with the other sites. Central Medusae Fossae 2 C in Figure 8(h) has the most tightly constrained age and erosion rate of all the sites. All six Central Medusae Fossae sites have well-defined $\beta$ values that are also all below 1000 nm yr$^{-1}$. However, other than Central Medusae Fossae 2 C, the five remaining sites have poorly constrained ages, with $2\sigma$ error regions that span around 1.2 to 3.4 Gyr, based on the marginalized CDFs. Of the six Central Medusae Fossae sites, the model had the most difficulty predicting Central Medusae Fossae 1 C (*1c*), in Figure 9(c). However, other than 1 C, and up to two bins per site, the Central Medusae Fossae distributions were in alignment with the data.

#### 3.4.2. Far East Medusae Fossae

For all three sites of Far East Medusae Fossae, the best fit of the model predicted young ages below 1 Gyr, but the $2\sigma$ region extends to older ages, such as up to 4.50 Gyr for Far East Medusae Fossae A (*5a*) and C (*5c*). The $\beta$ estimate for Far East Medusae Fossae A had the greatest uncertainty and the highest erosion rate estimate, at over 4000 nm yr$^{-1}$. Similarly, the age estimate for Far East Medusae Fossae A also had the greatest uncertainty, predicting a probability that included the entire 4.5 billion yr. The ultrafast erosion rate of Far East Medusae Fossae A (the two-parameter model's best fit of 4010 nm yr$^{-1}$, combined with its thickness (0.8 km, calculated in ArcGIS using an inverse distance-weighting interpolation), imply that the mountain will be destroyed in a geologically short time (200 Ma). Intermittent sediment inputs (e.g., ash or dust stone) may be preventing this from occurring. According to Figures 9(j)–(k), the model's prediction aligned well with the crater distributions of Far East Medusae Fossae B and C, though the model did not predict Far East Medusae Fossae A as accurately.

#### 3.4.3. East Medusae Fossae

The topographically high East Medusae Fossae B (*6b*) has far fewer craters than East Medusae Fossae A (*6a*). Though it is tempting to assume that B is much younger than A, our model finds broad and overlapping ages for A and B (the marginalized CDF results for age are $0.82^{+0.72}_{-0.24}$ Gyr and $0.90^{+1.99}_{-0.57}$ Gyr, respectively), with the difference in crater densities being explained by a much faster erosion rate for B than for A ($2590^{+310}_{-280}$ nm yr$^{-1}$ for B and $660^{+60}_{-50}$ nm yr$^{-1}$ for A).

#### 3.4.4. Zephyria Planum and Eastern Candor

The best-fit model predicts that Zephyria Planum (*2*) has a young age below 0.75 Gyr, while the upper $1\sigma$ bound on error extends up to 2.7 Gyr for Eastern Candor (*3*; Figures 8(d) and (e)). However, for Zephyria Planum, Figure 9(d) shows the best-fit line only falling slightly below the data. Eastern Candor, in particular, has very few craters, making it difficult for the model to predict an accurate best fit (Figure 9(e)). The thickness of Zephyria Planum (∼300 m, according to Kite et al. 2015), combined with its fast erosion rate, imply that it will be destroyed in a geologically short time (150 Myr).

#### 3.4.5. Upper Mount Sharp

Using the marginalized CDF, the best-fit age estimate is $0.95^{+0.90}_{-0.42}$ Gyr for Upper Mount Sharp (*7*). However, the crater data do not provide a useful maximum age for Upper Mount Sharp, which is instead provided by this unit's superposition on Lower Mount Sharp, which is of Hesperian age. Except for two bins, Figure 9(n) demonstrates that the best fit from the model aligned with the crater count data.

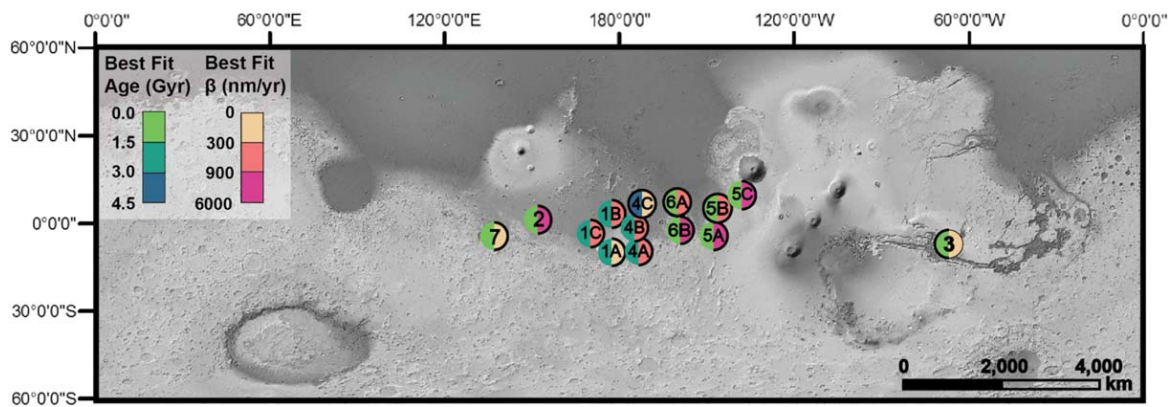

**Figure 10.** Map of Mars using MOLA topography and shaded relief (Smith et al. 2001), labeled with colored semicircles corresponding to the one-parameter model's best-fit age and best-fit β values (Figure 7) for each of the 14 study sites. Some circles are slightly offset from their geographic locations for the purposes of legibility. For the age and β semicircles with a thick black border, the site's 1σ range falls within the same increment as the best-fit value. For the semicircles without a border, the site's 1σ range falls outside the increment of the best-fit value. See Table 2 for details.

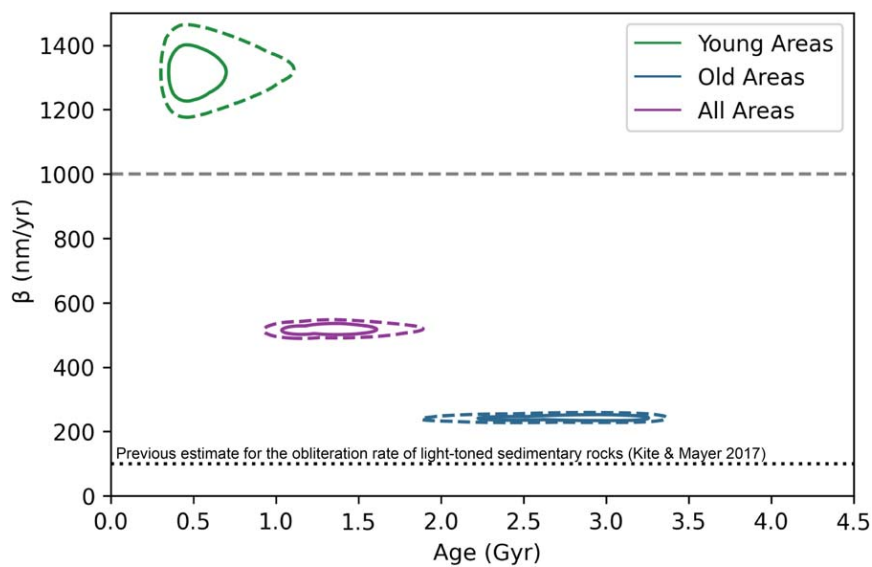

**Figure 11.** The green lines correspond to the fit of the model parameters to the data for the eight young sites (<1 Gyr) aggregated together; the blue lines correspond to the fit of the model parameters to the data for the six older sites (>1 Gyr) aggregated together; and the purple lines correspond to the fit of the model parameters to the data for all of the sites aggregated together. The solid lines represent the 1σ uncertainty sites and the dashed lines represent the 2σ uncertainty sites. The dashed gray line at 1000 nm yr⁻¹ indicates the minimum erosion rate needed for metastable equatorial water ice at depths shallow enough to be detected by GRS or NS. The black dotted line at $10^2$ nm yr⁻¹ indicates a previous estimate of the average of the typical obliteration rate for young light-toned sedimentary rock (Kite & Mayer 2017).

## 4. Discussion

### 4.1. Variations with Longitude

The map of the one-parameter model's best-fit age and erosion rates (Figure 10) demonstrates trends in age and erosion rate by location. Older ages greater than 1.5 Gyr dominate Central Medusae Fossae. East Medusae Fossae, Far East Medusae Fossae, Zephyria Planum, and Eastern Candor all have young best-fit ages, below 1 Gyr. Upper Mount Sharp, which is hypothesized to have similar material to the MFF (Zimbelman & Scheidt 2012; Lewis & Aharonson 2014; Edgett et al. 2020), also has a younger age, close to 0.9 Gyr. However, Zephyria Planum (2), Far East Medusae Fossae A (5a) and C (5c), East Medusae Fossae B (6b), and Upper Mount Sharp (7) all have 1σ uncertainties reaching into the next age increment (corresponding to the color bar in Figure 10) beyond the age increment of the best-fit age. Overall, most of Central Medusae Fossae has older ages and lower erosion rates, while most of

East Medusae Fossae, most of Far East Medusae Fossae, and Zephyria Planum have younger ages and higher erosion rates. Eastern Candor and Upper Mount Sharp are exceptions, as they have younger ages and relatively low erosion rates.

Figure 11 shows the two-parameter model predictions for the eight young sites aggregated together, the six older sites aggregated together, and all 14 sites aggregated together. The younger sites are predicted to have an age of 0.5 ± 0.2 Gyr. In contrast, while the aggregation of the older sites had a better constrained erosion rate of around 250 nm yr⁻¹, the age for the older sites was less well constrained, with a 1σ range spanning from 2.2 to 3.3 Gyr.

### 4.2. Model Uncertainties

The uncertainties in adopting the Hartmann isochrons (Hartmann 2005; Michael 2013) include crater identification, terrain selection, and secondary crater populations, as well as the application to Mars of a crater flux model that was originally based on radiometric age dating of lunar samples and comparisons to lunar crater counts (Hartmann 2005). Uncertainties relating to crater identification are inherent, due to the limited image resolution and crater-counting fatigue (Robbins et al. 2014). As mentioned in Section 2.1, the rolloff for our observed CSFD indicating survey incompleteness occurred at about 0.5 km, so we only included craters with diameters of 0.5 km and larger for our model.

For the terrain selection, other researchers might define the site boundaries differently from ours, especially sites *1*, *4*, *5*, and *6*, which have subregions within the boundaries based on the geological units defined in Tanaka et al. (2014). To analyze the effects of different terrain boundaries, we compared model estimates for age and erosion rate for sites *1*, *4*, *5*, and *6* before and after dividing them into subregions. The age and erosion rate estimates for the sites as a whole were broadly consistent with their subregions, with the ages toward the older side if one subregion was comparatively older with more craters than the other subregions.

To account for secondary craters, as mentioned in Section 2.1, apparent secondary craters (such as craters demonstrating high ellipticity, tight clustering, or with defining rays coming from the primary craters) were not included in the data set. Secondary craters that look like primary craters do not affect the CSFD model estimates, unless a large fraction of the craters come from a single event (Hartmann 2005). This is unlikely, since the MFF has wind-eroded rocks with different stratigraphic layers, so the craters were likely produced over a longer period of time.

The known problem of the overestimation of small craters by crater production functions (Hartmann 2005; Kite & Mayer 2017) is addressed by the inclusion of the erosion rate in our model (Figure 2). The probabilistic model does not take into account variable erosion rates, burial rates, and the obliteration of craters by impacts. Recent work has proposed a statistical model, based on crater size and depth frequency distribution, to determine obliteration rates as a function of time (Breton et al. 2022). While Breton et al. (2022) do not solve for age, their method does investigate how obliteration changes over time. Future work could expand upon our work that used a fixed erosion rate and age, and combine it with the changing erosion rate of Breton et al. (2022) by solving for the time-variable erosion rate and age of young sedimentary rock.

### 4.3. $\chi^2$ Test Comparison between the Two-parameter and One-parameter Models

To test whether our two-parameter model is preferable to a simpler one-parameter model, we performed minimum $\chi^2$ tests —an extension of $\chi^2$ goodness-of-fit tests (Wall & Jenkins 2012)—for the best fit based on the two-parameter model for age and $\beta$, the best fit along the *x*-axis for age, and the best fit along the *y*-axis for $\beta$, for all 14 sites grouped together (Table 3). All three tests were significant at the 99% confidence level, since the *p*-values were <0.01, so we can reject the null hypothesis that the crater distributions from the three models are the same as the crater count data.

The $\chi^2$ of 82.1 for the two-parameter model was the smallest. However, the $\chi^2$ exceeds the number of bins minus 1 then multiplied by 2, since $82.1 > 20$. As a result, we reject the null hypothesis that the two-parameter model's distribution was exactly the same as the crater count's distribution. The crater distribution predicted by the two-parameter model ("observed") was the most similar to the craters in the data ("expected"), in comparison to both one-parameter models. On the other hand, the $\chi^2$ values were much larger for both one-parameter models, which suggests that the values predicted by the one-parameter model are quite different from the data. The two-parameter model performs better than the one-parameter models when compared to the crater distribution data. For a distribution generated by a two-parameter model to better emulate the observed CSFD, we recommend future work that uses a time-varying erosion rate instead of a fixed erosion rate.

### 4.4. Implications for the Subsurface Ice Table and In Situ Resource Utilization

The subsurface ice table is the predicted location or depth of the boundary between icy ground and ice-free ground in the soil. The equatorial climate of Mars is extremely dry and relatively warm, which makes water ice unstable (Mellon & Jakosky 1995). As a result, the water ice at the equator can sublimate, forming water vapor that is then lost by diffusion through empty soil pore space, escaping to the dry atmosphere (ultimately going to the North Polar ice cap; Mellon & Jakosky 1995). With erosion rates of less than 1000 nm yr$^{-1}$, the ice can swiftly retreat to depths too deep to be detected by GRS or NS, which only probe around 1 m down (Boynton et al. 2002; Feldman et al. 2002; Mitrofanov et al. 2002; Feldman et al. 2004). By contrast, if the erosion rate is above 1000 nm yr$^{-1}$, the ice table can persist at shallow depths (∼1 m) in steady state because the erosion rate equals the ice table retreat rate under these conditions (Hudson et al. 2007). We obtain the erosion rate value of 1000 nm yr$^{-1}$ necessary for a shallow subsurface ice table from Figure 12 of Hudson et al. (2007). The right dashed line in this figure corresponds to retreat depth versus time for a higher value of the diffusion coefficient in the lag-forming case, with an initial barrier of 1 mm. The lag-forming case occurs when ice fills a low-porosity regolith, so that the removal of ice leaves behind a lag that increases in thickness (Hudson et al. 2007). From Figure 12 of Hudson et al. (2007), the right dashed line shows a retreat depth to 1 m in $10^6$ yr, which is equivalent to 1000 nm yr$^{-1}$ and corresponds to our white dashed line in Figure 8. We infer that if $\beta$ is greater than 1000 nm yr$^{-1}$, the depth of the buried ice can be less than 1 m, if there is any left.

The water ice would still be too deep for detection by cameras or IR spectroscopy, but it would be shallow enough to be detected by GRS/NS. As seen in Figure 12, Zephyria Planum (*2*), Far East Medusae Fossae A (*5a*), Far East Medusae Fossae C (*5c*), and East Medusae Fossae B (*6b*) are the four sites that have $\beta$ values

**Table 3**
Minimum $\chi^2$ Test Results for the Two-parameter and One-parameter Models

| Model Type | $\chi^2$ Test Statistic | *p*-value | Degrees of Freedom |
|---|---|---|---|
| Age (*x*-parameter) | 10700 | 0.0 | 9 |
| $\beta$ (one-parameter) | $5.10 \times 10^6$ | 0.0 | 9 |
| Age and $\beta$ (two-parameter) | 82.1 | $1.82 \times 10^{-14}$ | 8 |

**Note.** The degrees of freedom were calculated as $k - 1 - N$, where $k$ is the number of bins and $N$ is the number of parameters (Wall & Jenkins 2012). To perform the $\chi^2$ test, we combined the five largest crater bins that had <5 craters, so that $k = 11$ bins. This was necessary to ensure that the model-predicted number of craters was >5 craters for 80% of the bins.

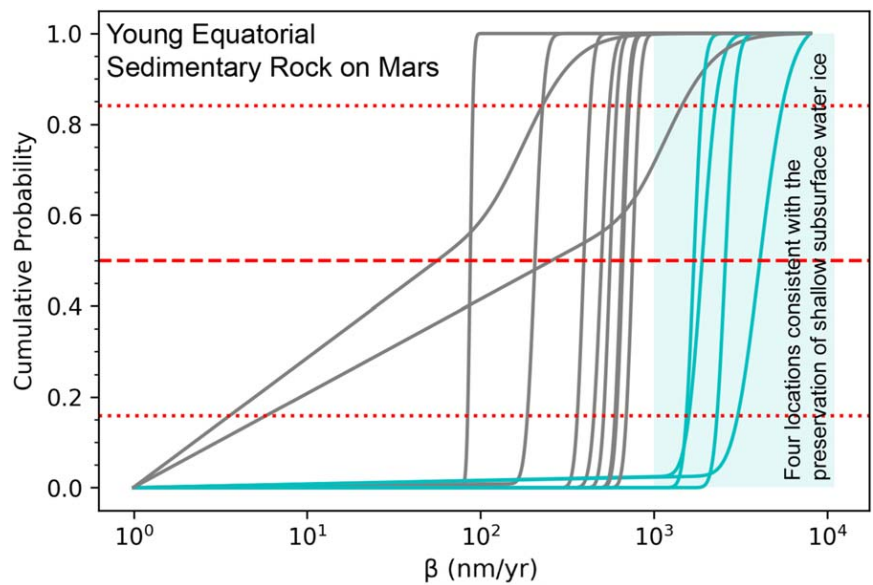

**Figure 12.** The four locations of the highest likelihood for water ice detection by GRS/NS are highlighted in teal. These include Zephyria Planum (*2*), Far East Medusae Fossae A and C (*5a* and *5c*), and East Medusae Fossae B (*6b*). The $\beta$ CDFs are the same as in Figure 7.

in the $1\sigma$ confidence region above 1000 nm yr$^{-1}$. All four sites are part of the young sites in Figure 11, which are expected to have erosion rates higher than 1000 nm yr$^{-1}$ when all eight young sites are aggregated together. Although the best-fit erosion rate based on the two-parameter model for Eastern Candor is predicted to be above 1000 nm yr$^{-1}$, the $1\sigma$ range is not confined to above 1000 nm yr$^{-1}$. The nine remaining sites, including Central Medusae Fossae (*1a–1c* and *4a–4c*), Far East Medusae Fossae B (*5b*), East Medusae Fossae A (*6a*), and Upper Mount Sharp (*7*), all have slower erosion rates. As a result, we expect Zephyria Planum, Far East Medusae Fossae A, Far East Medusae Fossae C, and East Medusae Fossae B to have the highest likelihood for water ice detection by GRS/NS.

Campbell et al. (2021) proposed a two-layer model for the MFF, with the top 300–600 m as a fine-grained, self-compacting layer, and the bottom 2 km as a thicker layer of possible water ice. If the self-compacting layer proposed by Campbell et al. (2021) did in fact extend to shallow depths (<10 m) as well, then it is unlikely that the material would be very rich in ice, though this is difficult to reconcile with the GRS and NS data (e.g., Wilson et al. 2018).

While two of the three sites in Campbell et al. (2021) are adjacent to, but do not overlap with, our study regions, for the region that does overlap at Zephyria Planum, Wilson et al. (2018) found a relatively high water equivalent hydrogen (WEH) content of >10% for the upper meter. Other western MFF locations, like Lucus Planum (referred to as Central Medusae Fossae in this paper) had >10% WEH as well, and Aeolis Planum had >40% WEH, which is too high to be explained by hydrated silicates, and instead strongly indicative of water ice (Wilson et al. 2018). Using the Fine Resolution Epithermal Neutron Detector neutron telescope on board the ExoMars Trace Gas Orbiter, Mitrofanov et al. (2022) recently reported high hydrogen abundances in central Valles Marineris that are consistent with the presence of water ice in the top 1 m of the subsurface. This motivates monitoring of the sites that we identified that are also in the equatorial region of Mars for small new impacts that might temporarily expose subsurface ice (Byrne et al. 2009; Dundas et al. 2014).

The potential presence of subsurface water ice at the equator of Mars has implications for in situ resource utilization (ISRU) as a source of liquid water and propellant. Liquid water is necessary for humans to survive and for life as we know it. While the poles of Mars have large amounts of water ice (Byrne 2009; Smith et al. 2009), this is less relevant to ISRU because early human missions are unlikely at those high latitudes (Starr & Muscatello 2020). Readily accessible water ice at minimal depths (on the meter scale) in the equatorial region of Mars would be invaluable for ISRU (Starr & Muscatello 2020; Bar-Cohen & Zacny 2021). In addition to liquid water, through the reaction $CO_2 + 2\,H_2O \rightarrow CH_4 + 2\,O_2$, water would also be useful for producing methane propellant. Producing propellant on Mars will be important in order to return a crew from the surface of Mars, since it is unlikely that the crew would be able to carry all of the propellant that they need with them from Earth (Ash et al. 1978; Rapp 2013; Heldmann et al. 2021). As a result, further investigations of potential subsurface ice tables in the equatorial and midlatitude regions of Mars will be key for future human missions (e.g., Morgan et al. 2021).

### 4.5. Implications for the Sedimentary Rock Cycle on Mars

The $\beta$ estimations from our model were consistent with previous work. Our estimated erosion rates for all sites combined (Figure 8(o)), Central Medusae Fossae 1 B and C (*1b* and *1c*; Figures 8(b)–(c)), and Central Medusae Fossae 2 A and B (*4a* and *4b*; Figures 8(f)–(g)) of 450–580 nm yr$^{-1}$ (Figure 8(o)) were all consistent with the Dunning (2019) estimated deflation rate of 400–600 nm yr$^{-1}$. Similarly, Grindrod & Warner (2014) state that erosion rates are 300–800 nm yr$^{-1}$, after converting horizontal retreat rates to vertical erosion, and Kite & Mayer (2017) found the crater obliteration rate of light-toned layered sedimentary rock to be $10^2$ nm yr$^{-1}$. The combined erosion rate from our model is $6.4 \times 10^{-4}$ km$^3$ yr, which is a 4.6 nm yr$^{-1}$ global equivalent. According to Ojha et al. (2018), the MFF may be the single largest source of dust on Mars, and they estimate a total volume loss of $3 \times 10^5$–$1.8 \times 10^6$ km$^3$ from the MFF, or a 2–12 m global layer of dust. By summing up all of the dust in the North Polar layered deposit (NPLD), the South Polar layered deposit (SPLD),

Arabia, Tharsis, Amazonis, Elysium, and other dusty regions, they estimate a global equivalent of around 3 m of dust (Ojha et al. 2018). By applying our erosion rate of $6.4 \times 10^{-4}$ km$^3$/yr, and treating the erosion rate as constant over long timescales, eroding the range of volumes estimated by Ojha et al. (2018) would require timescales of 0.47–2.8 Gyr. When we compare our estimated global equivalent erosion rate to the 3 m of dust from sources like the NPLD and SPLD, we would expect timescales of around 0.65 Gyr. Thus, our erosion rate estimates demonstrate that the MFF could in fact be a sustainable source of dust for the NPLD and SPLD.

According to Zimbelman & Scheidt (2012), the stratigraphically upper part of the lower member of the MFF is of early Amazonian age (Aml2), which is similar to layers near the top of Mount Sharp. This is in agreement with our collapsed one-parameter model results, which predict an age close to 0.9 Gyr for Upper Mount Sharp, which is similar in age to the young Amazonian ages that we predict for East and Far East Medusae Fossae. On the other hand, the other component of the lower member of the MFF analyzed by Zimbelman & Scheidt (2012) has an age near the Amazonian–Hesperian boundary (AHml1), which is consistent with the cratering ages of Mount Sharp as a whole (Thomson et al. 2011). These older ages are consistent with our one-parameter model estimations for Central Medusae Fossae. Previous work analyzing Mount Sharp within Gale Crater and the MFF has found that the dominant bedding scales are similar but not identical for the two sites ($0.22 \pm 0.02$ m$^{-1}$ at Gale versus $0.35 \pm 0.04$ m$^{-1}$ and $0.29 \pm 0.06$ m$^{-1}$ for the MFF), and suggested a possible analogous formation link (Lewis & Aharonson 2014). After the initial formation of the MFF, Kerber & Head (2010) predict sediment recycling at Aeolis Mons, and, in fact, extraformational sediment recycling, the process of sedimentary rock being repeatedly buried then exposed by erosion, is known to occur at the top of Aeolis Mons (Edgett et al. 2020). Due to the similarities between Aeolis Mons and the MFF, it is possible that similar processes have occurred at the MFF.

The purpose of this study was to provide constraints on the absolute ages and erosion rates of the MFF and not its formation mechanism. However, our results are consistent with the leading hypothesis of initial emplacement in the Hesperian, with later reworking in the Amazonian (e.g., Kerber & Head 2010; Kerber et al. 2011). In a recycling hypothesis, the formation age that we calculate represents the time of the most recent deposition. The Central Medusae Fossae (Lucus Planum) was predicted to have overall older ages (1.5–4.5 Gyr) and lower erosion rates (<650 nm yr$^{-1}$), while Zephyria Planum, Far East Medusae Fossae, East Medusae Fossae, Eastern Candor, and Aeolis Mons were predicted to have younger ages with large erosion rates (>650 nm yr$^{-1}$; Figure 10). This suggests that Zephyria Planum, East Medusae Fossae, and Far East Medusae Fossae most likely have young Amazonian ages, consistent with the most sediment recycling, while reworking plays less of a role for the older terrains of Central Medusae Fossae. While Kerber & Head (2010) also predicted sediment recycling at Aeolis Mons, we predict both Aeolis Mons and Easter Candor to have young ages, with relatively lower rates of erosion, perhaps indicating slower reworking than Zephyria Planum, East Medusae Fossae, and Far East Medusae Fossae.

In the future, this model may also be applied to additional sedimentary locations on Mars. In particular, the model can provide more precise age and erosion rate estimations for regions that contain larger amounts of craters. For these 14 sites, the model has been able to estimate a much more precise erosion rate than age, as seen in Figure 10. In addition, the geologic context of a location may provide further constraints, to enable the model to find either the erosion rate or age. For example, if the geologic context of a location suggests very minimal erosion, then one could use this constraint and the one-parameter model to find the allowed error regions for age. As a result, future work should also focus on methods for better constraining age.

## 5. Conclusions

There are few constraints on the absolute formation age of young equatorial sedimentary rock on Mars because of the wind erosion that obliterates the impact craters. To work around this problem, we constructed a model of the crater size distribution function that allows both the age and erosion rate $\beta$ to vary. We applied the model to 14 equatorial sites. Central Medusae Fossae 1 C was the only site that was clearly constrained to an absolute model age of 2.98–3.51 Gyr by the two-parameter model, which is consistent with the Kerber & Head (2010) inference of the MFF forming during the Hesperian. Other sites were individually less well constrained. On the other hand, the erosion rate had much better constraints, and nine of 14 sites had $\beta$ values that were constrained below 1000 nm yr$^{-1}$. In addition, 12 of 14 sites had $\beta$ values that were larger than 10 nm yr$^{-1}$, with only the lower $1\sigma$ error bound falling below 10 nm yr$^{-1}$ for two sites, suggesting that the 14 sites had non-negligible erosion.

We identified four sites as having the highest potential for the preservation of subsurface ice at shallow depths: Zephyria Planum, Far East Medusae Fossae A and C, and East Medusae Fossae B (Section 4.4). Although these four sites were among the sites with the broadest erosion rate fits, these four sites still had $2\sigma$ certainty for an erosion rate above 1000 nm yr$^{-1}$ (and even $3\sigma$ certainty for Far East Medusae Fossae C and East Medusae Fossae B). As a result, these four locations have the highest likelihood of water ice detection by GRS/NS. Thus, we recommend further investigations of these four sites, due to their potential implications for ISRU.

These four sites have best-fit formation ages within the past 1 billion yr (Figures 7 and 8), consistent with Amazonian age estimates (Scott & Tanaka 1986; Greeley & Guest 1987). This suggests that—but does not require—deposition and/or sediment-recycling events and aqueous cementation occurred within the last 1 billion yr. However, the upper $2\sigma$ error bounds permit older ages as well. For all six Central Medusae Fossae sites, deposition occurred more than 1 billion yr ago, according to Figures 7 and 8. From the global map of the one-parameter model's best-fit values for age and erosion rate, Figure 10 shows common trends between the sites. While Central Medusae Fossae had older ages (>1.40 Gyr) and lower erosion rates (<650 nm yr$^{-1}$), East Medusae Fossae, Far East Medusae Fossae, and Zephyria Planum all had younger ages (<1 Gyr) and higher erosion rates (>740 nm yr$^{-1}$). Eastern Candor and Aeolis Mons were exceptions, as they were predicted to have young ages and relatively low erosion rates.

We acknowledge Daniel Fabrycky and Michael Foote for their valuable insights on an early draft. We are very grateful for constructive reviews from Lujendra Ojha and an anonymous reviewer. In addition, we also thank the HiWish program. E.S.K. acknowledges funding from NASA (80NSSC20K0144).

## Appendix

For Python scripts of the model figures on GitHub, please access the following link: https://github.com/AstroAnLi/young_sedimentary_rock_mars. Please email the primary author with any questions.

## ORCID iDs

An Y. Li  https://orcid.org/0000-0002-6151-5214
Edwin S. Kite  https://orcid.org/0000-0002-1426-1186